
\def\title{Correlation Functions of the XXZ model for $\Delta<-1$}
\def\author{$Michio Jimbo{}^a$, Kei Miki${}^b$,
Tetsuji Miwa${}^c$ and Atsushi Nakayashiki${}^d$\cr
{}\cr
${}^a$
Department of Mathematics,Faculty of Science,\cr
Kyoto University, Kyoto 606, Japan \cr
{}\cr
${}^b$
Department of Mathematical Science,
Faculty of Engineering Science,\cr
Osaka University, Toyonaka, Osaka 560, Japan \cr
{}\cr
${}^c$
Research Institute for Mathematical Sciences,\cr
Kyoto University, Kyoto 606, Japan\cr
{}\cr
${}^d$
The Graduate School of Science and Technology, \cr
Kobe University, Rokkodai, Kobe 657, Japan \cr
}
\par
\def\rhead{Correlation Functions of the XXZ model for $\Delta<-1$}
\def\lhead{M. Jimbo et al.}

\hsize=5.3truein
\vsize=7.8truein
\baselineskip=10pt
\font\twelvebf=cmbx12
\nopagenumbers
\headline={\ifnum\pageno=1 \hss\sl RIMS\hss \else \hdline\fi}
\def\hdline{\ifodd\pageno\rightheadline \else\leftheadline\fi}
\def\rightheadline{\tenrm\hfil{\it \rhead}\hfil\folio}
\def\leftheadline{\tenrm\folio\hfil{\it \lhead}\hfil}
\voffset=2\baselineskip
\vglue 2cm
%
%
\centerline{\twelvebf
\vbox{
\halign{\hfil # \hfil\cr
\title\crcr}}}
%
%
\bigskip
%
%
\centerline{\tenrm
\vbox{
\halign{\hfil#\hfil\cr
\author\crcr}}}
%
%
\vglue 1cm
%
%
%
\def\address #1
{\vglue 0.5cm
\halign{\quad\it##\hfil\cr
#1\crcr}}
{\it Dedicated to Professor Chen Ning Yang on the occasion of his seventieth
birthday}
\vskip 1.5cm

\par
{\narrower\bigskip{\noindent\bf Abstract.\quad}
A new approach to the correlation functions is presented
for the XXZ model in the anti-ferroelectric regime.
The method is based on the recent realization of the
quantum affine symmetry using vertex operators.
With the aid of a boson representation for the latter,
an integral formula is found for correlation functions
of arbitrary local operators.
As a special case it reproduces the spontaneous staggered polarization
obtained earlier by Baxter.
\bigskip}
\par

\message{Cross-reference macros, B. Davies, version 9 May 1992.}

\catcode`@=11


\newif\if@xrf\@xrffalse   
\def\l@bel #1 #2 #3>>{\expandafter\gdef\csname @@#1#2\endcsname{#3}}
\immediate\newread\xrffile
\def\xrf@n#1#2{\expandafter\expandafter\expandafter
\csname immediate\endcsname\csname #1\endcsname\xrffile#2}
\def\xrf@@n{\if@xrf\relax\else%
  \expandafter\xrf@n{openin}{ = \jobname.xrf}\relax%
  \ifeof\xrffile%
    \message{ no file \jobname.xrf - run again for correct forward references
}%
  \else%
    \expandafter\xrf@n{closein}{}\relax%
    \catcode`@=11 \input\jobname.xrf \catcode`@=12%
  \fi\global\@xrftrue%
  \expandafter\expandafter\csname immediate\endcsname%
  \csname  newwrite\endcsname\xrffile%
  \expandafter\xrf@n{openout}{ = \jobname.xrf}\relax\fi}


\newcount\t@g

\def\order#1{%
  \expandafter\expandafter\csname newcount\endcsname
  \csname t@g#1\endcsname\csname t@g#1\endcsname=0
  \expandafter\expandafter\csname newcount\endcsname
  \csname t@ghd#1\endcsname\csname t@ghd#1\endcsname=0

  \expandafter\def\csname #1\endcsname##1{\xrf@@n\csname n@#1\endcsname##1:>}

  \expandafter\def\csname n@#1\endcsname##1:##2>%
    {\def\n@xt{##1}\ifx\n@xt\empty%
     \expandafter\csname n@@#1\endcsname##1:##2:>
     \else\def\n@xt{##2}\ifx\n@xt\empty%
     \expandafter\csname n@@#1\endcsname\unp@ck##1 >:##2:>\else%
     \expandafter\csname n@@#1\endcsname\unp@ck##1 >:##2>\fi\fi}

  \expandafter\def\csname n@@#1\endcsname##1:##2:>%
    {\edef\t@g{\csname t@g#1\endcsname}\edef\t@@ghd{\csname t@ghd#1\endcsname}%
     \ifnum\t@@ghd=\t@ghd\else\global\t@@ghd=\number\t@ghd\global\t@g=0\fi%
     \ifunc@lled{@#1}{##1}\global\advance\t@g by 1%
       {\def\n@xt{##1}\ifx\n@xt\empty%
       \else\writ@new{#1}{##1}{\pret@g\t@ghead\number\t@g}\expandafter%
       \xdef\csname @#1##1\endcsname{\pret@g\t@ghead\number\t@g}\fi}%
       {\pret@g\t@ghead\number\t@g}%
     \else\def\n@xt{##1}%
       \w@rnmess#1,\n@xt>\csname @#1##1\endcsname%
     \fi##2}%

  \expandafter\def\csname ref#1\endcsname##1{\xrf@@n%
     \expandafter\each@rg\csname #1cite\endcsname{##1}}

  \expandafter\def\csname #1cite\endcsname##1:##2,%
    {\def\n@xt{##2}\ifx\n@xt\empty%
     \csname #1cit@\endcsname##1:##2:,\else%
       \csname #1cit@\endcsname##1:##2,\fi}

  \expandafter\def\csname #1cit@\endcsname##1:##2:,%
    {\def\n@xt{\unp@ck##1 >}\ifunc@lled{@#1}{\n@xt}%
      {\expandafter\ifx\csname @@#1\n@xt\endcsname\relax%
       \und@fmess#1,\n@xt>>>\n@xt<<%
       \else\csname @@#1\n@xt\endcsname##2\fi}%
     \else\csname @#1\n@xt\endcsname##2%
     \fi}}


\def\each@rg#1#2{{\let\thecsname=#1\expandafter\first@rg#2,\end,}}
\def\first@rg#1,{\callr@nge{#1}\apply@rg}
\def\apply@rg#1,{\ifx\end#1\let\n@xt=\relax%
\else,\callr@nge{#1}\let\n@xt=\apply@rg\fi\n@xt}

\def\callr@nge#1{\calldor@nge#1-\end-}
\def\callr@ngeat#1\end-{#1}
\def\calldor@nge#1-#2-{\ifx\end#2\thecsname#1:,%
  \else\thecsname#1:,\hbox{\rm--}\thecsname#2:,\callr@ngeat\fi}


\def\unp@ck#1 #2>{\unp@@k#1@> @>>}
\def\unp@@k#1 #2>>{\ifx#2@\@np@@k#1\else\@np@@k#1@> \unp@@k#2>>\fi}
\def\@np@@k#1#2#3>{\ifx#2@\@@np@@k#1>\else\@@np@@k#1>\@np@@k#2#3>\fi}
\def\@@np@@k#1>{\ifcat#1\alpha\expandafter\@@np@@@k\string#1>\else#1\fi}
\def\@@np@@@k#1#2>{@#2}


\def\writ@new#1#2#3{\xrf@@n\immediate\write\xrffile
  {\noexpand\l@bel #1 #2 {#3}>>}}


\def\ifunc@lled#1#2{\expandafter\ifx\csname #1#2\endcsname\relax}
\def\und@fmess#1#2,#3>{\ifx#1@%
  \message{ ** error - eqn label >>#3<< is undefined ** }\else
  \message{ ** error - #1#2 label >>#3<< is undefined ** }\fi}
\def\w@rnmess#1#2,#3>{\ifx#1@%
  \message{ Warning - duplicate eqn label >>#3<< }\else
  \message{ Warning - duplicate #1#2 label >>#3<< }\fi}


\def\t@ghead{}
\newcount\t@ghd\t@ghd=0
\def\taghead#1{\gdef\t@ghead{#1}\global\advance\t@ghd by 1}


\order{@qn}


\let\eqno@@=\eqno
\def\eqno(#1){\xrf@@n\eqno@@\hbox{{\rm(}$\@qn{#1}${\rm)}}}

\let\leqno@@=\leqno
\def\leqno(#1){\xrf@@n\leqno@@\hbox{{\rm(}$\@qn{#1}${\rm)}}}

\def\refeq#1{\xrf@@n{{\rm(}$\ref@qn{#1}${\rm)}}}


\def\eqalignno#1{\xrf@@n\displ@y \tabskip=\centering
  \halign to\displaywidth{\hfil$\displaystyle{##}$\tabskip=0pt
   &$\displaystyle{{}##}$\hfil\tabskip=\centering
   &\llap{$\eqaln@##$}\tabskip=0pt\crcr
   #1\crcr}}

\def\leqalignno#1{\xrf@@n\displ@y \tabskip=\centering
  \halign to\displaywidth{\hfil$\displaystyle{##}$\tabskip=0pt
   &$\displaystyle{{}##}$\hfil\tabskip=\centering
    &\kern-\displaywidth\rlap{$\eqaln@##$}\tabskip\displaywidth\crcr
   #1\crcr}}

\def\eqaln@#1#2{\relax\ifcat#1(\expandafter\eqno@\else\fi#1#2}
\def\eqno@(#1){\xrf@@n\hbox{{\rm(}$\@qn{#1}${\rm)}}}


\def\n@@me#1#2>{#2}
\def\numberby#1{\xrf@@n
  \ifx\s@ction\undefined\else
  \expandafter\let\csname\s@@ve\endcsname=\s@ction\fi
  \ifx\subs@ction\undefined\else
  \expandafter\let\csname\subs@@ve\endcsname=\subs@ction\fi
  \numb@rby#1,>#1>}
\def\numb@rby#1,#2>#3>{\def\n@xt{#1}\ifx\n@xt\empty\taghead{}\else
  \def\n@xt{#2}\ifx\n@xt\empty\n@by#3>\else\n@@by#3>\fi\fi}
\def\n@by#1>{\ifx\s@cno\undefined\expandafter\expandafter
  \csname newcount\endcsname\csname s@cno\endcsname
  \csname s@cno\endcsname=0\else\s@cno=0\fi
  \xdef\s@@ve{\expandafter\n@@me\string#1>}
  \let\s@ction=#1\def#1{\global\advance\s@cno by 1
  \taghead{\number\s@cno.}\s@ction}}
\def\n@@by#1,#2>{\ifx\s@cno\undefined\expandafter\expandafter
  \csname newcount\endcsname\csname s@cno\endcsname
  \csname s@cno\endcsname=0\else\s@cno=0\fi
  \ifx\subs@cst\undefined\expandafter\expandafter
  \csname newcount\endcsname\csname subs@cst\endcsname
  \csname subs@cst\endcsname=0\else\subs@cst=0\fi
  \ifx\subs@cno\undefined\expandafter\expandafter
  \csname newcount\endcsname\csname subs@cno\endcsname
  \csname subs@cno\endcsname=0\else\subs@cno=0\fi
  \xdef\s@@ve{\expandafter\n@@me\string#1>}
  \let\s@ction=#1\def#1{\global\advance\s@cno by 1
  \global\subs@cno=\subs@cst
  \taghead{\number\s@cno.}\s@ction}
  \xdef\subs@@ve{\expandafter\n@@me\string#2>}
  \let\subs@ction=#2\def#2{\global\advance\subs@cno by 1
  \taghead{\number\s@cno.\number\subs@cno.}\subs@ction}}


\def\numberfrom#1{\ifx\s@cno\undefined\else\n@mberfrom#1,>\fi}
\def\n@mberfrom#1,#2>{\def\n@xt{#2}%
  \ifx\n@xt\empty\n@@f#1>\else\n@@@f#1,#2>\fi}
\def\n@@f#1>{\s@cno=#1\advance\s@cno by -1}
\def\n@@@f#1,#2,>{\s@cno=#1\advance\s@cno by -1%
  \subs@cst=#2\advance\subs@cst by -1}


\def\pret@g{}
\def\prefixby#1{\gdef\pret@g{#1}}



\newcount\r@fcount\r@fcount=0
\newcount\r@fcurr
\newcount\r@fone
\newcount\r@ftwo
\newif\ifc@te\c@tefalse
\newif\ifr@feat

\def\refto#1{{\rm[}\def\s@p{}\refn@te#1>>\refc@te#1>>{\rm]}}

\def\refn@te#1>>{\refn@@te#1,>>}

\def\refn@@te#1,#2>>{\r@fnote{\expandafter\unp@ck\str@pbl#1 >> >}%
   \def\n@xt{#2}\ifx\n@xt\empty\else\refn@@te#2>>\fi}

\def\refc@te#1>>{\r@fcurr=0\r@featfalse\def\s@ve{}%
  {\loop\ifnum\r@fcurr<\r@fcount\advance\r@fcurr by 1\c@tefalse%
   \expandafter\refc@@te\number\r@fcurr>>#1,>>%
   \ifc@te\expandafter\refe@t\number\r@fcurr>>\fi\repeat\s@ve}}

\def\refc@@te#1>>#2,#3>>{\def\n@xt{\expandafter\unp@ck\str@pbl#2 >> >}%
   \expandafter\refc@@@te\csname r@f\n@xt\endcsname>>#1>>%
   \def\n@xt{#3}\ifx\n@xt\empty\else\refc@@te#1>>#3>>\fi}

\def\refc@@@te#1>>#2>>{\ifnum#2=#1\relax\c@tetrue\fi}

\def\refe@t#1>>{\ifr@feat\ifnum\r@fone=\r@ftwo\res@cond#1>>%
   \else\reth@rd#1>>\fi\else\r@feattrue\ref@rst#1>>\fi}

\def\ref@rst#1>>{\r@feattrue\r@fone=#1\r@ftwo=#1%
   \s@p\expandafter\relax\number\r@fone}%

\def\res@cond#1>>{\advance\r@ftwo by 1\def\n@xt{#1}%
   \expandafter\ifnum\n@xt=\number\r@ftwo%
   \edef\s@ve{,\expandafter\relax\number\r@ftwo}\else,\ref@rst#1>>\fi}%

\def\reth@rd#1>>{\advance\r@ftwo by 1\def\n@xt{#1}%
   \expandafter\ifnum\n@xt=\number\r@ftwo%
   \edef\s@ve{--\expandafter\relax\number\r@ftwo}\def\s@p{,}\else%
   \s@ve\def\s@ve{}\ref@rst#1>>\fi}%

\def\r@fnote#1%
  {\ifunc@lled{r@f}{#1}\global\advance\r@fcount by 1%
   \expandafter\xdef\csname r@f#1\endcsname{\number\r@fcount}%
   \expandafter\gdef\csname r@ftext\number\r@fcount\endcsname%
   {\message{ Reference #1 to be supplied }%
   Reference $#1$ to be supplied\par}\fi}

\def\str@pbl#1 #2>>{#1#2}


\def\refis#1 #2\par{\def\n@xt{\unp@ck#1 >}\r@fis\n@xt>>#2>>}
\def\r@fis#1>>#2>>{\ifunc@lled{r@f}{#1}\else
   \expandafter\gdef\csname r@ftext\csname
r@f#1\endcsname\endcsname{#2\par}\fi}


\def\listreferences{\global\r@fcurr=0%
  {\loop\ifnum\r@fcurr<\r@fcount\global\advance\r@fcurr by 1%
   \numr@f\number\r@fcurr>>\csname r@ftext\number\r@fcurr\endcsname>>%
   \repeat}}

\def\numr@f#1>>#2>>{\vbox{\noindent\hang\hangindent=30truept%
   {\hbox to 30truept{\rm[#1]\hfill}}#2}\smallskip\par}


\def\printlabels{\global\@xrftrue\def\s@me##1{$##1$}
  \def\@qn##1{##1}\def\refeq##1{{\rm(}$##1${\rm)}}\refbylabel
  \def\numberby##1{\relax}\def\numberfrom##1{\relax}
  \def\listreferences{\relax}\def\referencefile{\relax}
  \def\order##1{\expandafter\let\csname ##1\endcsname=\s@me
                \expandafter\let\csname ref##1\endcsname=\s@me}}

\def\refbylabel{\def\refto##1{[$##1$]}%
  \def\refis##1 ##2\par{\numr@f$##1$>>##2>>}\def\numr@f##1>>##2>>%
  {\noindent\hang\hangindent=30truept{{\rm[##1]}\ }##2\par}}


\def\beginsection#1\par{\vskip0pt plus.3\vsize\penalty-250
  \vskip0pt plus -.3\vsize\bigskip\vskip\parskip
  \leftline{\bf#1}\nobreak\smallskip\noindent}

\catcode`@=12

\def\C{{\bf C}}

\def\Q{{\bf Q}}
\def\Z{{\bf Z}}

\def\F{{\cal F}}

\def\la{\lambda}
\def\La{\Lambda}

\def\Hom{\hbox{Hom}}
\def\End{\hbox{End}}
\def\id{\hbox{id}}
\def\tr{\hbox{tr}}

\def\slt{\goth{sl}_2}
\def\slth{\widehat{\goth{sl}}_2\hskip 2pt}

\def\goto#1{{\buildrel #1 \over \longrightarrow}}
\def\br#1{\langle #1 \rangle}

\def\brak#1#2{\langle #1|#2\rangle}

\def\vac{|\hbox{vac}\rangle}
\def\dvac{\langle \hbox{vac}|}
%
%



%
\font\germ=eufm10
\def\goth#1{\hbox{\germ #1}}
%
%
\def\Figure(#1|#2|#3)
{\midinsert
\vskip #2
\hsize 9cm
\raggedright
\noindent
{\bf Figure #1\quad} #3
\endinsert}
%
%
\def\Table #1. \size #2 \caption #3
{\midinsert
\vskip #2
\hsize 7cm
\raggedright
\noindent
{\bf Table #1.} #3
\endinsert}
%
\def\sectiontitle#1\par{\vskip0pt plus.1\vsize\penalty-250
 \vskip0pt plus-.1\vsize\bigskip\vskip\parskip
 \message{#1}\leftline{\bf#1}\nobreak\vglue 5pt}
\def\qed{\hbox{${\vcenter{\vbox{
    \hrule height 0.4pt\hbox{\vrule width 0.4pt height 6pt
    \kern5pt\vrule width 0.4pt}\hrule height 0.4pt}}}$}}
\def\subsec(#1|#2){\medskip\noindent{\it #1}\hskip8pt{\it #2}\quad}
\def\eq#1\endeq
{$$\eqalignno{#1}$$}
\def\leq#1\endeq
{$$\leqalignno{#1}$$}
%
%
\def\qbox#1{\quad\hbox{#1}\quad}
%
%

%
%

%
%

%
%

%
%

%
%

%
%
\def\Definition#1.#2{\smallskip\noindent {\sl Definition #1.#2\quad}}
%
%
\def\subsec(#1|#2){\medskip\noindent#1\hskip8pt{\sl #2}\quad}
%
%
%
%
\def\abstract#1\endabstract{
\bigskip
\itemitem{{}}
{\bf Abstract.}
\quad
#1
\bigskip
}
%
%
%
\def\sec(#1){Sect.\hskip2pt#1}

\order{prop}
\numberby{\beginsection}\numberfrom{0}


%
%
%
%
%

\def\s{\sigma}
\def\o{\otimes}
\def\XXZ{{X\hskip-2pt X\hskip-2pt Z}}
\def\qq{F}

\beginsection \S 0. Introduction

In this article we consider the
one-dimensional infinite spin-chain
\eq
&H_{\XXZ}=-{1\over 2}\sum_{k=-\infty}^\infty
\bigl( \s^x_{k+1}\s^x_k+\s^y_{k+1}\s^y_k
+\Delta\s^z_{k+1}\s^z_k \bigr) &(Ham)\cr
\endeq
known classically as the XXZ model.
We shall limit ourselves strictly to
the anti-ferroelectric regime $\Delta<-1$.
Our aim is to find an exact expression for the spin correlation functions
$\br{\s^{\gamma_1}_{i_1} \cdots \s^{\gamma_n}_{i_n}}$, where
$\s^\gamma_i=\s^x_i,\s^y_i,\s^z_i$ and
$\br{\cdot }=\dvac \cdot \vac$ denotes the ground state average.
Here the ground state $\vac$ means one of the two ground states in the
anti-ferroelectric regime. In Baxter's paper \refto{Bax}
they are denoted by $|\pm\rangle$.
The present paper is based entirely on the framework of the
recent work \refto{DFJMN},
which discribes the realization of the quantum affine symmetry
for \refeq{Ham} using the $q$-deformed vertex operators.
Let us recall below the contents of \refto{DFJMN} which are
relevant to the subsequent discussions.

The Hamiltonian \refeq{Ham} is an operator acting on the
infinite tensor product  $V^{\o\infty}=\cdots V\!\o\!V\!\o\!V\!\o \cdots$
of the two dimensional space $V=\C^2$.
The space $V^{\o\infty}$ admits also
an action of the quantum affine algebra $U=U_q(\slth)$
via the iterated coproduct.
A na\"{\i}ve computation shows that the algebra $U$ provides an exact symmetry
for \refeq{Ham}.
Namely if $\Delta=(q+q^{-1})/2$, then $[H_{\XXZ},U']=0$ where
$U'=U'_q(\slth)$ denotes the subalgebra
`without the grading operator $d$' \refto{DFJMN},
while $d$ plays the role of the boost operator.
However the actions of \refeq{Ham} and $U$ are both defined only formally, and
the issue is how to extract the theory free from
the difficulties of divergence.

The basic idea in \refto{DFJMN} is to replace the formal object
$V^{\o \infty}$ by the level $0$ $U$-module
\eq
&\F_{\la,\mu}=V(\la)\widehat{\o} V(\mu)^{*a}\simeq \Hom(V(\mu), V(\la)),
&(Hilb)
\endeq
where $V(\la)$ ($\la=\La_0,\La_1$) denotes the level $1$ highest weight
$U$-module, and $V(\la)^{*a}$ signifies its dual (the superfix indicates that
the $U$-module structure is given via the antipode $a$).
The choice $\la=\mu$ (resp. $\la\neq \mu$) is responsible for the even
(resp. odd) particle sector. $V(\La_0)$ (resp. $V(\La_1)$) means
that we are working in the boundary condition $\s^z_{2k}=1,\s^z_{2k+1}=-1$
(resp. $\s^z_{2k}=-1,\s^z_{2k+1}=1$) for $k>\hskip-2pt>0$,
and $V(\La_0)^{*a}$ (resp. $V(\La_1)^{*a}$) means
the same thing for $k<\hskip-2pt<0$.
The tensor product should be completed in the $q$-adic sense to allow
for infinite sums (see \refto{DFJMN} for a precise treatment).
To make contact with the na\"{\i}ve picture of $V^{\o \infty}$,
one utilizes the embedding of
$V(\la)$ into the half infinite tensor product
$\cdots V\!\o\!V\!\o\!V$ \refto{FM}.
This is supplied by iterating the vertex operators
\eq
&\Phi_\la^{\mu V}:V(\la) \goto{} V(\mu)\o V. &(VOI)\cr
\endeq
(It is conjectured \refto{DFJMN}
that there is a unique normalization of \refeq{VOI}
which makes the infinite iteration convergent.)
Similarly $V(\mu)^{*a}$ embeds to the other half infinite tensor product
$V\!\o\!V\!\o\!V\!\o \cdots$, giving altogether the embedding
\eq
&\F_{\la,\mu}~ \goto{}~ \cdots V\o V\o V \o \cdots. &(embd) \cr
\endeq
Eq. \refeq{embd} provides the principle
of interpreting the notions defined in the picture $V^{\o \infty}$
by pulling them back to $\F_{\la,\mu}$.
The translation operator $T$ is the first such example.
(The Hamiltonian \refeq{Ham} itself
is defined in terms of $T$ and the grading operator $d$;
see \refto{DFJMN}.)
In this paper we follow the same principle to formulate
the local operators $\s^\gamma_k$ as acting on $\F_{\la,\mu}$.

In the language of \refeq{Hilb}
the ground state (vacuum) vector is given by the identity element of
$\F_{\la,\la}=\Hom(V(\la),V(\la))$.
The inner product of two vectors $f,g\in \F_{\la,\la}$ is given
as the trace
\eq
&\brak{f}{g}={\tr_{V(\la)}(q^{-2\rho}fg)
\over \tr_{V(\la)}(q^{-2\rho})} &(inner)
\endeq
where $\rho=\La_0+\La_1$ (we have changed the normalization from
\refto{DFJMN}, see \sec(1) below).
Thus the correlation functions
$\dvac \s^{\gamma_1}_{i_1} \cdots \s^{\gamma_n}_{i_n} \vac$
can be expressed as the trace of products of the
vertex operators \refeq{VOI}.

To perform the evaluation of these traces,
we invoke the bosonization method.
The realization of the $q$-deformed currents on level one modules
was done in \refto{FrJ} using (ordinary) bosons.
In the same spirit we derive the formulas for the vertex operators \refeq{VOI}
in terms of bosons.
This leads to an explicit formula for the correlators
in terms of certain integrals of meromorphic functions.
We verify that in the simplest case of the one-point function $\br{\s^z_k}$
this formula reproduces the known result for the spontaneous staggered
polarization due to Baxter \refto{Bax}. Jacques Perk noted a strong similarity
between our formula and the formula for the Ising model
spin-spin correlation functions given in \refto{Zinv,AYP}.

The bosonization of the vertex operators also enables us to compute the
$n$-point functions discussed in Sect. 6.8 of \refto{DFJMN}.
They are the matrix elements (not the trace)
of the product of the vertex operators with respect to
the highest weight vectors.
The conjectural formula (6.39) of \refto{DFJMN}
is thus proved. In this paper, we do not go into
details on this matter.

The plan of the paper is as follows.
In \sec(1) we formulate the local operators and their correlators
using the scheme mentioned above.
In our algebraic formulation the algebra $U$, intertwiners, etc.
are a priori defined over the base field $\qq=\Q(q)$
($q$ an indeterminate, see the remark at the end of Sect. 2);
we shall see however that
the resulting formulas are meaningful for complex values of $q$ with $|q|<1$.
In \sec(2) we outline the bosonization of the vertex operators.
The basic ingredients are the Drinfeld realization of the algebra $U'$
and the result of \refto{FrJ}, which we recall briefly.
Unlike the classical case ($q=1$) only one of the components $\Phi_-(z)$
of \refeq{VOI} has the exponential form, while the other one does not
but is
given as the $q$-commutator of $\Phi_-(z)$ with a generator $f_1$ of $U$.
Sect. 3 is devoted to the formula for the correlators.

\def\qbox#1{\quad\hbox{#1}}
\def\slth{\widehat{\goth{sl}}(2)\hskip 1pt}
\def\slt{\goth{sl}(2)\hskip 1pt}
\def\u{U_q\bigl(\slth \bigr)}
\def\Up{U'_q\bigl(\slth\bigr)}
\def\L{\Lambda}
\def\VO(#1,#2){{\tilde \Phi}^{#1}_{#2}}
\def\h(#1){|u_{#1}\rangle}
\def\cL{{\cal{L}}}

\def\e{\varepsilon}

\beginsection \S 1. Vacuum expectation values of local operators

Let us recall some notations of \refto{DFJMN}.
We denote by $V=\qq v_+\oplus \qq v_-$ the 2-dimensional vector space
on which the Pauli matrices $\sigma^x,\sigma^y,\sigma^z$ act.
We consider $V$ as the 2-dimensional $\Up$-module.
The $\XXZ$-Hamiltonian formally acts on the infinite tensor product
$V^{\o\infty}$ of $V$.
We label the components of the tensor product by integers $k \in \Z$
from right ($k\rightarrow -\infty$) to left ($+\infty\leftarrow k$).
In our mathematical scheme, we replace
the semi-infinite tensor product of the components $k\ge 1$
by the irreducible highest weight $\u$-module $V(\L_i)\ (i=0,1)$
Let us define the action of the local operators on $V(\L_i)$.
Let $L\in\End(\underbrace{V\o\cdots\o V}_n)$.
The operator $L$ naturally acts on the tensor product of
the components $n\ge k\ge 1$ of $V^{\o\infty}$.
We want to interprete this action as one in $\End\bigl(V(\L_i)\bigr)$.
For this purpose, we use the vertex operators
\refto{DFJMN}: they are the intertwiners
\eq
&\VO(\mu V,\lambda)(z):V(\lambda)\rightarrow V(\mu)\o V_z,
\quad \VO(\mu V,\lambda)(z)(v)=
\VO(\mu V,\lambda+)(z)(v)\o v_+
+\VO(\mu V,\lambda-)(z)(v)\o v_- \quad &(cmpt)\cr
&\VO(\mu,\lambda V)(z):V(\lambda)\o V_z\rightarrow V(\mu),
\quad \VO(\mu,\lambda V)(z)(v\o v_\pm)=
\VO(\mu,\lambda V){}_\pm(z)(v),\cr
\endeq
where $\lambda=\L_0,\mu=\L_1$ or $\lambda=\L_1,\mu=\L_0$, and
$V_z=V\o F[z,z^{-1}]$ signifies
the $\u$-module associated to $V$ with the
spectral parameter $z$ (see \refto{DFJMN}, \sec(6)).
Precisely speaking, we consider the vertex operators as the generating
functions of their Fourier components in terms of the spectral parameter $z$.
They are normalized as
\eq
&\VO(\La_1 V,\La_0)(z)(\h(\La_0))
=\h(\La_1)\o v_-+\cdots,\quad
\VO(\La_0 V,\La_1)(z)(\h(\La_1))
=\h(\La_0)\o v_++\cdots,&(Norm)\cr
&\VO(\La_1,\La_0 V){}_+(z)=\VO(\La_1 V,\La_0-)(z/q^2),\qquad
\VO(\La_1,\La_0 V){}_-(z)=-q^{-1}\VO(\La_1 V,\La_0+)(z/q^2),\cr
&\VO(\La_0,\La_1 V){}_+(z)=-q\VO(\La_0 V,\La_1-)(z/q^2),\qquad
\VO(\La_0,\La_1 V){}_-(z)=\VO(\La_0 V,\La_1+)(z/q^2),\cr
&{(q^2;q^4)_\infty\over(q^4;q^4)_\infty}
\VO(\lambda,\mu V)(z) \VO(\mu V,\lambda)(z)
=\id_{V(\lambda)}.\cr
\endeq
Here $\h(\La_i)$ denotes the highest weight vector of $V(\La_i)$.
We have used the standard notation
$(z;p)_\infty=\prod_{j=0}^{\infty}(1-z p^j)$.

Set $\L_n=\L_{n-2}$ for simplicity of notation.
Given an $L$ as above, we define the operator
$\cL=\rho^{(i)}_{z_n,\ldots,z_1}(L)\in\End\bigl(V(\L_i)\bigr)$ by
\eq
\cL=&{(q^2;q^4)_\infty^n\over(q^4;q^4)_\infty^n}\
\VO(\L_i,\L_{i+1}V)(z_1)\circ\cdots\circ\VO(\L_{i+n-1},\L_{i+n}V)(z_n)\cr
&\circ\bigl({\rm id}_{V(\L_{i+n})}\o L\bigr)\circ
\VO(\L_{i+n}V,\L_{i+n-1})(z_n)\circ\cdots\circ\VO(\L_{i+1}V,\L_i)(z_1). \cr
\endeq

In this paper, we use the following convention
(different from that of \refto{DFJMN})
for the invariant bilinear form on
$P=\Z\L_0\oplus\Z\L_1\oplus\Z\delta$:
\eq
&(\L_0,\L_0)=0,(\L_0,\alpha_1)=0,(\L_0,\delta)=1,\cr
&(\alpha_1,\alpha_1)=2,(\alpha_1,\delta)=0,(\delta,\delta)=0.\cr
\endeq
Note that $\L_1=\L_0+\alpha_1/2$, $\delta=\alpha_0+\alpha_1$.
We set $\rho=\L_0+\L_1$ as usual, and also $\alpha=\alpha_1$
for simplicity. We identify $P^*=\Z h_0\oplus \Z h_1\oplus \Z d$
as a subset of $P$ via $(\ ,\ )$. We have $\rho=2d+\alpha_1/2$.

The vacuum expectation value
$\langle L\rangle^{(i)}_{z_n,\ldots,z_1}$
of the local operator $L$ is given by
\eq
&\langle L\rangle^{(i)}_{z_n,\ldots,z_1}
={\tr_{V(\L_i)}\bigl(q^{-2\rho}\rho^{(i)}_{z_n,\ldots,z_1}(L)\bigr)
\over \tr_{V(\L_i)}\bigl(q^{-2\rho}\bigr) }.&(tr)\cr
\endeq
This is a consequence of the formula for the invariant inner product
in the space $V(\L_i)\o V(\L_i)^{*a}$ \refto{DFJMN}.
For the $\XXZ$ model correlator we specialize the spectral parameters to
$z_1=\cdots=z_n$.
We expect that the formula unspecialized
would give the equal-row vertical arrow correlator
for the inhomogeneous six-vertex model with $z_k$ being the trigonometric
spectral parameter of the $k$-th vertical line.
Here we use the usual language of vertex models on the 2-dimensional
square lattice, in which the fluctuation variables are described
as arrows sitting on vertical or horizontal edges.
Note the following selection rule, which is specific to the six vertex model:
$\langle L\rangle^{(i)}_{z_n,\ldots,z_1}=0$
for $L=E_{\e'_n\e_n}\o\cdots\o E_{\e'_1\e_1}$
($E_{ij}$ is a matrix unit)
such that $\e_1+\cdots+\e_n
\not=\e'_1+\cdots+\e'_n$.

\def\ga{\gamma}
\def\del{\delta}

\def\o{\otimes}
\def\e{\hbox{e}}

\beginsection \S 2. Bosonization

Let us first recall Drinfeld's realization of the
quantum affine algebra $U'=\Up$ \refto{Dri}.
It is an associative algebra
generated by the letters
$\{x_k^{\pm}\mid k\in{\bf Z}\}$,
$\{a_l \mid l\in{\bf Z}_{\neq0}\}$,
$\ga^{\pm 1/2}$ and $K$,
satisfying the following defining relations.
\eq
&\ga^{\pm 1/2}\in\hbox{ the center of the algebra},
\cr
&[a_k,a_l]=\del_{k+l,0}{1\over k}[2k]{\ga^k-\ga^{-k}\over q-q^{-1}},
\quad [a_k,K]=0,
\cr
&Kx_k^{\pm}K^{-1}=q^{\pm2}x_k^{\pm},
\cr
&[a_k,x_l^{\pm}]=\pm {1\over k}[2k]\ga^{\mp |k|/2}x_{k+l}^{\pm},
\cr
&x_{k+1}^{\pm}x_l^{\pm}-q^{\pm2}x_{l}^{\pm}x_{k+1}^{\pm}=
q^{\pm2}x_{k}^{\pm}x_{l+1}^{\pm}-x_{l+1}^{\pm}x_k^{\pm},
\cr
&[x_k^{+},x_l^{-}]={1\over q-q^{-1}}(\ga^{(k-l)/2}\psi_{k+l}
-\ga^{(l-k)/2}\varphi_{k+l}),\cr
\endeq
where $[n]=(q^n-q^{-n})/(q-q^{-1})$ and
$\{\psi_r,\varphi_s \mid r\in{\bf Z}_{\geq 0},s\in{\bf Z}_{\le  0 } \}$
are related to $\{a_l\mid l\in{\bf Z}_{\not=0}\}$  by
\eq
\sum_{k=0}^\infty\psi_kz^{-k}&
=K\exp\Bigl((q-q^{-1})\sum_{k=1}^\infty
a_kz^{-k}\Bigr),\cr
\sum_{k=0}^\infty\varphi_{-k}z^{k}&
=K^{-1}\exp\Bigl(-(q-q^{-1})\sum_{k=1}^\infty
a_{-k}z^{k}\Bigr).\cr
\endeq

The standard Chevalley generators $\{e_i,f_i,t_i\}$
are given by the identification
\eq
&t_0=\ga K^{-1},~
t_1= K,~
e_1= x_0^+,~
f_1= x_0^-,~
e_0t_1= x_1^-,~
t_1^{-1}f_0=x_{-1}^+.
\endeq
We use the coproduct
\eq
\Delta(e_i)&=e_i\o1+t_i\o e_i,\cr
\Delta(f_i)&=f_i\o t_i^{-1}+1\o f_i,\cr
\Delta(t_i)&=t_i\o t_i.\cr
\endeq

Next recall the construction of the level one irreducible
integrable highest weight representations of $U'$
in terms of bosons \refto{FrJ}.
Let $P={\bf Z}{\alpha\over2}$, $Q={\bf Z}\alpha$
be the weight/root lattice of $\slt$, and let
$\qq[P]$, $\qq[Q]$ be their group algebras.
The basis elements of $\qq[P]$ are written
multiplicatively as $\e^{n\alpha}$ $(n\in {1\over2}{\bf Z})$.
As an $\qq[Q]$-module, $\qq[P]=\qq[P]_0\oplus \qq[P]_1$
where $\qq[P]_i=\qq[Q]\e^{i\alpha/2}$.
Introduce a $U'$-module structure on the space
$W=\qq[a_{-1},a_{-2},\cdots]\o\qq[P]$
in the following way.
First let $\{a_k\}$, $\e^\beta$ and $\partial_\alpha$ act on $W$ as
\eq
a_k&=\hbox{the left multiplication by $a_k\o1$ for $k<0$}, \cr
   &=[a_k,\cdot]\o1 \qquad \hbox{for $k>0$}, \cr
\e^{\beta_1}(f\o\e^{\beta_2})&=f\o\e^{\beta_1+\beta_2}, \cr
\partial_\alpha(f\o\e^{\beta})&=(\alpha,\beta)f\o\e^{\beta}. \cr
\endeq
We let also
\eq
&K=1\o q^{\partial_\alpha}, \qquad \gamma=q\o \id.
\endeq
The actions of the generators $\{x^{\pm}_n\}$ are given through the
generating functions $X^{\pm}(z)=\sum_{n\in{\bf Z}}x_n^{\pm}z^{-n-1}$
as follows.
\eq
X^+(z)&=
\exp\bigl(\sum_{n=1}^\infty{a_{-n}\over[n]}q^{-n/2}z^n\bigr)
\exp\bigl(-\sum_{n=1}^\infty{a_{n}\over[n]}q^{-n/2}z^{-n}\bigr)
\e^\alpha z^{\partial_\alpha},
\cr
X^{-}(z)&=
\exp\bigl(-\sum_{n=1}^\infty{a_{-n}\over[n]}q^{n/2}z^n\bigr)
\exp\bigl(\sum_{n=1}^\infty{a_{n}\over[n]}q^{n/2}z^{-n}\bigr)
\e^{-\alpha} z^{-\partial_\alpha}.
\cr
\endeq
With these actions $W$ becomes a $U'$-module.
The submodules $\qq[a_{-1},a_{-2},\cdots]\o\qq[P]_i$ are
isomorphic to the irreducible highest weight modules $V(\Lambda_i)$
with the highest weight vectors $u_{\Lambda_0}=1\o1$ and
$u_{\Lambda_1}=1\o\e^{\alpha/2}$.
The grading operator $d$ is introduced by
\eq
-d(a_{-i_1}^{n_1}\cdots a_{-i_r}^{n_r}\o\e^\beta)&
=\bigl(\sum_{j=1}^rn_ji_j+{(\beta,\beta)\over2}-{(\La_i,\La_i)\over2}
\bigr)
(a_{-i_1}^{n_1}\cdots a_{-i_r}^{n_r}\o\e^\beta).
\endeq

We remark that the trace of $p^{-d}X^{+}(z)X^{-}(z)$ on
$V(\La_0)$ is calculated in \refto{Jg}.
We shall describe the vertex operators
in terms of the representation constructed above,
and then compute a similar trace for them (see Sect. 3).

Let the components
${\tilde \Phi}_{\pm}(z)={\tilde \Phi}_{\la\pm}^{\mu V}(z)$ be defined as
in \refeq{cmpt}.
{}From the condition that it intertwines the action of $x_0^{-}$ we find
\eq
&{\tilde \Phi}_{+}(z)=[{\tilde \Phi}_{-}(z),x_0^{-}]_q,&(phip)
\endeq
where $[X,Y]_q=XY-qYX$.
The intertwining relations with the Chevalley generators determine
${\tilde \Phi}_{\pm}(z)$ uniquely, but it is not easy to get
the expression of the vertex operators from them only. We better
use the relations with Drinfeld's generators $a_n$.
We lack the formulas for the coproduct of them, in general.
However we have the following partial information
which will suffice for our purpose \refto{CP}.

\proclaim Proposition \prop{cop}.
For $k\geq0$ and $l>0$ we have
\eq
\Delta(x_k^{+})&=x_k^{+}\o\ga^k+\ga^{2k}K\o x_k^{+}+
\sum_{i=0}^{k-1}\ga^{(k+3i)/2}\psi_{k-i}\o\ga^{k-i}x_i^{+}
{}~~\bmod N_{-}\o N_{+}^2,
\cr
\Delta(x_{-l}^{+})&=x_{-l}^{+}\o\ga^{-l}+K^{-1}\o x_{-l}^{+}+
\sum_{i=1}^{l-1}\ga^{(l-i)/2}\varphi_{-l+i}\o\ga^{-l+i}x_{-i}^{+}
{}~~\bmod N_{-}\o N_{+}^2,
\cr
\Delta(a_l)&=a_l\o\ga^{l/2}+\ga^{3l/2}\o a_l
{}~~\bmod N_{-}\o N_{+},
\cr
\Delta(a_{-l})&=a_{-l}\o\ga^{-3l/2}+\ga^{-l/2}\o a_{-l}
{}~~\bmod N_{-}\o N_{+}.
\endeq
Here $N_{\pm}$ and $N_{\pm}^2$ are left
$F[\ga^\pm,\psi_r,\varphi_s|r,-s\in{\bf Z}_{\geq0}]$-modules
 generated by $\{x_m^{\pm}|k\in {\bf Z} \}$ and
$\{x_m^{\pm}x_n^{\pm}|m,n\in {\bf Z}\}$ respectively.

By using Proposition \refprop{cop}
and noting that $N_\pm v_\pm=0,$ $N_{+}v_{-}\subset F[z,z^{-1}]v_{+}$,
we get the exact relations
\eq
[a_k,{\tilde \Phi}_{-}(z)]&
=q^{7k/2}{[k]\over k}z^k{\tilde \Phi}_{-}(z)\qbox{$k>0$},
\cr
[a_{-k},{\tilde \Phi}_{-}(z)]&
=q^{-5k/2}{[k]\over k}z^{-k}{\tilde \Phi}_{-}(z)\qbox{$k>0$},
\cr
[{\tilde \Phi}_{-}(z),X^{+}(w)]&=0,\quad
t_1{\tilde \Phi}_{-}(z)t_1^{-1}=q{\tilde \Phi}_{-}(z).
\cr
\endeq
These conditions along with the normalization \refeq{Norm}
determine the form of ${\tilde \Phi}_{-}(z)$ completely.
Explicitly we have, on $V(\Lambda_i)$ $(i=0,1)$,
\eq
&{\tilde \Phi}_{-}(z)=
\exp\bigl(
\sum_{n=1}^\infty{a_{-n}\over[2n]}q^{7n/2}z^n\bigr)
\exp\bigl(
-\sum_{n=1}^\infty{a_{n}\over[2n]}q^{-5n/2}z^{-n}\bigr)
\e^{\alpha/2}(-q^3z)^{(\partial_\alpha+i)/2}.  &(phim)
\endeq
It can be checked directly that the operators given by \refeq{phip,phim}
enjoy the correct intertwining properties.

Let us make a comment on the base field $F=\Q(q)$. In the above expression
of the bosonization of the vertex operators, the square root of $q$ appears.
But, it can be absorbed in the boson oscillators if we change the
definition of their commutation relations.
Note also that the combination ${(\partial_\alpha+i)/2}$ in the power of
$-q^3z$ produces only an integer power acting on $V(\La_i)$.
In any event, the appearance of $q^{1/2}$ is only superficial,
and our theory is free from the ambiguity in the choice of the square root.



\def\qbox#1{\quad\hbox{#1}}
\def\XXZ{{X\hskip-2pt X\hskip-2pt Z}}
\def\L{\Lambda}
\def\Q{{\bf Q}}
\def\cL{{\cal{L}}}

\def\lo{\langle{\rm 0}|}
\def\ro{|{\rm 0}\rangle}
\def\tr{{\rm tr}}
\def\Z{{\bf Z}}
\def\o{\otimes}
\def\e{\varepsilon}
\def\O{{\cal O}}

\def\al{\alpha}
\def\pr#1{(#1)_{\infty}}

\def\z{\zeta}
\def\vl{\hskip0.7mm |\hskip0.7mm}
\def\mu#1{E_{\e'_#1\,\e_#1}}
\def\sig(#1,#2){\sigma_#2^{#1}}
\def\tPhi{{\tilde \Phi}}
\def\VO(#1,#2){{\tilde \Phi}^{#1}_{#2}}
\def\QV(#1,#2,#3){\VO(\L_{#3}V,\L_{#2}\e_#1)(z_#1)}
\def\QVV(#1,#2,#3){\VO(\L_{#3},\L_{#2}V\e_#1')(z_#1)}
\def\C{{\cal C}}
\def\qq{F}
\def\F{{\cal F}}
\def\zz{\bar z}
\def\ee{\bar \eta}

\beginsection \S 3. Calculation of correlators

As explained in Sect.1,  for the evaluation of the correlators it is necessary
to calculate the trace of the product of the vertex operators.
Using the boson realization
described in Sect.2, we shall treat the trace of the form
$\tr(x^{-d}y^\al\O)$  where $x$, $y$ are complex parameters with $|x|<1$.
Since $q^{-2\rho}=q^{-4d-\al}$, the choice $x=q^4$, $y=q^{-1}$ will be
relevant to the correlators.
The calculation  is simplified by the technique
of  Clavelli and Shapiro (\refto{cl}, Appendix C).
Their prescription is as follows. Introduce a copy of bosons $\{b_n\}$
satisfying
$[a_m,b_n]=0$ and the same commutation relations as the $a_n$.
Let
\eq
&{\tilde a}_n={a_n\over 1-x^n}+b_{-n} \quad(n> 0),\quad
a_{n}+{b_{-n} \over x^n-1} \quad(n< 0).
\endeq
For a linear operator $\O=\O(\{a_n\})$ on the Fock space
$\F_a=F[a_{-1},a_{-2},\cdots]$, let ${\tilde \O}=\O({\{\tilde a_n}\})$ be the
operator on $\F_a\o\F_b$ $(\F_b=F[b_{-1},b_{-2},\cdots]$)
 obtained by substituting
$\tilde a_n$ for $a_n$.
We have then
\eq
&\tr_{\F_a}\bigl(x^{-{\bar d}}\O\bigr)={\lo{\tilde \O}\ro \over
\prod_{n=1}^{\infty}(1-x^n)},\cr
\endeq
where
${\bar d}=-\sum_{n>0}{n^2\over[n][2n]}a_{-n}a_n$ and
$\lo{\tilde \O}\ro$ denotes the usual expectation value with respect to
the Fock vacuum $\ro=1\o 1\in \F_a\o\F_b$, $\lo{\rm 0}\rangle =1$.

In order to apply their method to our case, we need the following.
Set
\eq
&f(z)=\pr{zx;x}\pr{q^2x/z;x},\quad
g(z)=\prod_{n=1}^{\infty}{\pr{zq^2 x^n;q^4}\over \pr{zq^4 x^n;q^4}}.\cr
\endeq
Let us denote the operators $X^-(z)$ and $\tPhi_\pm(z)$ with ${\tilde a}_n$
substituted
for $a_n$ by $J(z)$ and $\phi_\pm(z)$.
Then the explicit expressions  of $J(z)$ and $\phi_{-}(z)$ (after the normal
ordering) are
\eq
J(z)&=f(1)\exp\bigl(-\sum_{n=1}^\infty{q^{n/2}\over[n]}
(z^{n} a_{-n}-z^{-n}b_{-n})\bigr)\cr
&\times \exp\bigl(\sum_{n=1}^\infty{q^{{n/2}}\over[n]}
{z^{-n} a_{n}-(xz)^{n}b_{n}\over 1-x^n} \bigr)
e^{-\alpha} z^{-\partial_\alpha},\cr
\phi_{-}(z)&=g(1)\exp\bigl(\sum_{n=1}^\infty{
q^{7n/2}z^{n} a_{-n}-q^{-5n/2}z^{-n}b_{-n}\over [2n]}\bigr)\cr
&\times\exp\bigl(-\sum_{n=1}^\infty
{q^{-5n/2}z^{-n} a_{n}-q^{7n/2}(xz)^{n}b_{n}\over [2n]( 1-x^n)}
 \bigr) e^{{\alpha/2}}(-q^3z)^{(\partial_\alpha+i)/2}
\hbox{ on }V(\L_i).
\endeq
They satisfy the following relations,
\eq
J(\xi_1)J(\xi_2)&=:J(\xi_1)J(\xi_2):
\biggl(1-{\xi_2\over\xi_1}\biggr)\biggl(1-{q^2 \xi_2\over\xi_1}\biggr)
f\biggl({\xi_2\over\xi_1}\biggr)f\biggl({\xi_1\over\xi_2}\biggr),\cr
\phi_{-}(z_1)\phi_{-}(z_2)&=:\phi_{-}(z_1)\phi_{-}(z_2):
g\biggl({z_1\over z_2}\biggr)g\biggl({z_2\over z_1x}\biggr),
\cr
J(\xi)\phi_{-}(z)&=:J(\xi)\phi_{-}(z):{1\over (1-q^2 w^{-1})f(w)},\cr
\phi_{-}(z)J(\xi)&=:J(\xi)\phi_{-}(z):{-q^{-1}w\over (1-w)f(w)},\cr
\phi_{+}(z)=(q-q^{-1})&\oint_{q^2< |w|< 1}
{d\xi\over 2\pi i }:J(\xi)\phi_{-}(z):{w\over (1-w)(1-q^2 w^{-1})f(w)}.
\cr
\endeq
Here $w=\xi/q^2 z$ and $:\cdots:$ denotes the normal ordering with respect to
the bosons $a_n$ and $b_n$ (We do not normal order the operators
$e^{n\al}$ and $\partial_\al$).
In the above equations, all factors of the form $1/(1-z)$
should be understood as $\sum_{n=0}^{\infty}z^n$.
This fact specifies the contour for the expression $\phi_{+}$.
Using the formulas above, \refeq{tr} is evaluated as follows.

Set
\eq
&P_{\e_n',\cdots,\e_1'}^{\e_n,\cdots,\e_1}(z_n,\cdots,z_1\vl x,y\vl i)
={\pr{q^2;q^4}^n\over \pr{q^4;q^4}^n}\cr
&\times{\tr_{V(\L_i)}
\bigl(x^{-d}y^{\al}\QVV(1,i+1,i)\cdots\QVV(n,i+n,i+n-1)
\QV(n,i+n-1,i+n)\cdots\QV(1,i,i+1)\bigr)\over
\tr_{V(\L_i)}(x^{-d}y^\al)}.&(gen)\cr
\endeq
We have
\eq
&\langle L\rangle^{(i)}_{z_n,\ldots,z_1}=
P_{\e_n',\cdots,\e_1'}^{\e_n,\cdots,\e_1}(z_n,\cdots,z_1\vl q^4,q^{-1}\vl i).
\cr
\endeq
for $L=E_{\e'_n\e_n}\o\cdots\o E_{\e'_1\e_1}$
(see the remark at the end of Sect. 1).

Introduce the following notations
\eq
&A=\{a_1,\cdots,a_s\}=\{j\vl\e_j'=-1\},\quad
B=\{b_1,\cdots,b_t\}=\{j\vl\e_j=+1\},\cr
&(s+t=n,\quad a_i<a_j,\quad b_i<b_j \qbox{for } i<j),\cr
&h(z)=\pr{q^2z;x}\pr{xz;x}\pr{q^2z^{-1};x}\pr{xz^{-1};x}.\cr
\endeq
We prepare the integration variables $\xi_a$ $(a\in A)$, $\zeta_b$ $(b\in B)$
and set
$\eta_j=\xi_{a_j}$ $(1\le j\le s)$, $=\z_{b_{n+1-j}}$ $(s<j\le n)$,
$\ee=\prod_j\eta_j$ and $\zz=\prod_j z_j$.
Then we have
\eq
&P_{\e_n',\cdots,\e_1'}^{\e_n,\cdots,\e_1}(z_n,\cdots,z_1\vl x,y\vl i)\cr
&=(-1)^t q^{\sum_{a\in A}a+\sum_{b\in B}b-n(n+1)/2}
\prod_{a\in A}\oint_{\C_a} {d\xi_a \over 2\pi i (\xi_a-z_a)}
\prod_{b\in B}\oint_{\C_b} {d\z_b \over 2\pi i
(\z_b-z_b)}\cr
&\times\prod_{a \in A}\prod_{a<j\le n}{z_j-q^2 \xi_a \over z_j-\xi_a}
\prod_{b \in B}\prod_{b<j\le n}{\z_b-q^2 z_j \over
\z_b-z_j}
\prod_{j<k}{\eta_k-\eta_j \over \eta_k-q^2\eta_j}\cr
&\times{h(1)^n \prod_{j<k}h(z_j/z_k)h(\eta_j/\eta_k) \over
\prod_{j,k}h(\eta_j/z_k)}
{\sum_{m\in\Z+i/2}(\zz/\ee)^{2m}y^{2m}x^{m^2-i/4}
\over\pr{x;x}\tr_{V(\L_i)} (x^{-d}y^{\al})}.&(P)
\cr
\endeq
Note that the last factor of the above equation can be rewritten into
\eq
&\Bigl({\zz\over\ee}\Bigr)^i
{\pr{-(y\zz/\ee)^2x^{1+i};x^2}\pr{-(\ee/y\zz)^2 x^{1-i};x^2}\over
\pr{-y^2x^{1+i};x^2}\pr{-y^{-2}x^{1-i};x^2}}.\cr
\endeq
The contours of integration should be chosen as follows.
Both $\C_a$ and $\C_b$ are anti-clockwise, and the $z_i\ (1\le i\le n)$
lie inside of $\C_a$ and outside of $\C_b$. Other relevant poles
with small multipliers $q,x$ (e.g., $q^2z_i$) are inside, and those
with large multipliers are outside of the contours $\C_a$ and $\C_b$.
By specializing \refeq{P} to the case $x=q^4$ and $y=q^{-1}$, we obtain
the integral formula for the correlators.
The case $n=1$ gives the spontaneous staggered polarization
$P_0=P^+_+(1\vl1)-P^-_-(1\vl1)$ \refto{Bax}.
Note that  $P^\e_\e(1\vl i)=P^{-\e}_{-\e}(1\vl 1-i)$.
By using \refeq{P} for $P^+_+(1\vl 1)$ and $P^+_+(1\vl 0)$
we get $P_0=\pr{q^2;q^2}^2/\pr{-q^2;q^2}^2$.

%
\catcode`@=11
\newif\ifs@p
\def\refjl#1#2#3#4%
  {#1\def\l@st{#1}\ifx\l@st\empty\s@pfalse\else\s@ptrue\fi%
   \def\l@st{#2}\ifx\l@st\empty\else%
   \ifs@p, \fi{\frenchspacing\sl#2}\s@ptrue\fi%
   \def\l@st{#3}\ifx\l@st\empty\else\ifs@p, \fi{\bf#3}\s@ptrue\fi%
   \def\l@st{#4}\ifx\l@st\empty\else\ifs@p, \fi#4\s@ptrue\fi%
   \ifs@p.\fi\hfill\penalty-9000}
\def\refbk#1#2#3%
  {#1\def\l@st{#1}\ifx\l@st\empty\s@pfalse\else\s@ptrue\fi%
   \def\l@st{#2}\ifx\l@st\empty\else%
   \ifs@p, \fi{\frenchspacing\sl#2}\s@ptrue\fi%
   \def\l@st{#3}\ifx\l@st\empty\else\ifs@p, \fi#3\s@ptrue\fi%
   \ifs@p.\fi\hfill\penalty-9000}
\catcode`@=12
%
%

\def\CMP{Commun. Math. Phys.}

\def\IJMPA{Int. J. Mod. Phys. A}

\def\JSP{J. Stat. Phys.}

\def\NPB{Nucl. Phys. B}

\def\RIMS{RIMS preprint}

\par
\bigskip\noindent
{\it Acknowledgement.}\quad
We wish to thank N. H. Jing for sending his preprint
\refto{Jg}, and H. Ooguri and J. H. H. Perk for discussions.
We are grateful to Brian Davies and Omar Foda
for their collaborations in our previous works, on which
is based the present paper.

\par
\bigskip\noindent{\bf References}\medskip
\par
\refis{Bax} \refjl
{Baxter R J,
Spontaneous staggered polarization of the $F$ model}
{\JSP}{9}{(1973) 145--182}
\par
\refis{DFJMN} \refjl
{Davies B, Foda O, Jimbo M, Miwa T and Nakayashiki A,
Diagonalization of the XXZ Hamiltonian by vertex operators}
{\RIMS}{873}{(1992)}
\par
\refis{FM} \refjl
{Foda O and Miwa T,
Corner transfer matrices and quantum affine algebras}
{\IJMPA}{7}{supplement 1A (1992) 279--302}
\par
\refis{FrJ} \refjl
{Frenkel I B and Jing N H,
Vertex representations of quantum affine algebras}
{Proc. Nat'l. Acad. Sci. USA}{85}{(1988) 9373--9377}
\par
\refis{CP} \refjl
{Chari V and Pressley A,
Quantum affine algebras}
{\CMP}{142}{(1991) 261--283}
\par
\refis{Dri} \refjl
{Drinfeld V G,
A new realization of Yangians and quantum affine algebras}
{Soviet Math. Doklady}{36}{(1988) 212--216}
\par
\refis{cl} \refjl
{Clavelli L and Shapiro J A,
Pomeron Factorization in General Dual Models}
{\NPB}{57}{(1973) 490--535}
\par
\refis{Zinv} \refjl
{Baxter R J,
Solvable eight-vertex model on an arbitrary planar lattice}
{Phil. Trans. Roy. Soc.}{A289}{(1978) 315--346}
\par
\refis{AYP} \refjl
{Au-Yang H and Perk J H H,
Critical Correlations in a $Z$-invariant Inhomogeneous Ising Model}
{Physica} {144A} {(1987) 44--104}
\par
\refis{Jg} \refbk
{Jing N H,
On a trace of $q$ analog vertex operators}
{in Quantum Groups, Spring Workshop on Quantum Groups
Argonne National Laboratory 16 April--11 May 1990,
eds. T. Curtright, D. Fairlie and C. Zachos}
{}{World Scientific, Singapore 1992}
\par
\listreferences
\par

\end